**Towards Unifying Resilience and Sustainability for Transportation Infrastructure Systems: Conceptual Framework, Critical Indicators, and Research Needs**

**H M Imran Kays**
Ph.D. Student
School of Civil Engineering & Environmental Science
University of Oklahoma
202 W. Boyd St., Norman, OK 73019-1024
Email: imran_kays@ce.mist.ac.bd

**Arif Mohaimin Sadri, Ph.D.**
Incoming Assistant Professor
School of Civil Engineering & Environmental Science
University of Oklahoma
202 W. Boyd St., Norman, OK 73019-1024
E-mail: sadri@ou.edu
(Corresponding Author)

Word Count: 01 tables + 11 figures + 5671 words = 5921 word equivalents

**ABSTRACT**

Sustainability aspects of transportation infrastructure systems primarily focus on system performance based on environmental, social, and economic impacts. In contrast, resilience aspects demonstrate the ability to withstand external shocks i.e. robustness as well as to recover from the loss of functionality due to such disruptions i.e. rapidity. Therefore, sustainability and resilience are two key aspects which should be given adequate attention during the planning, design, construction, operations, and maintenance phases of any civil infrastructure system. As both concepts are equally important to sustain an infrastructure for a longer duration, their concurrent assessments within a unified framework are highly desirable. While there has been a recent focus towards solving this dilemma, review of existing studies revealed the lack of such unifying frameworks that can quantify sustainability and resilience indicators to simultaneously assess system performance. Moreover, a single decision or performance indicator could reinforce one and undermine another. As such, this study proposed a forward-looking unification framework, where the sustainability and resilience of transportation infrastructure systems can be analyzed simultaneously. In this regard, the proposed unifying framework is explained using seven critical indicators including emission, speed, temperature, energy consumption, delay, mobility, and accessibility. This study also investigates the interdependencies, relationships, and tradeoffs between sustainability and resilience based on these indices. While some indices would help the system to attain both at the same time, they are compromised in some cases. Finally, the study identifies immediate research needs as well as the ones in the long-term.

**Keywords:** Transportation Infrastructure Systems, Resilience Indices, Sustainability Indices, Unification

**INTRODUCTION AND MOTIVATION**

Sustainability has been a key consideration for transportation infrastructure throughout its lifespan for long-lasting development. The Brundtland Commission first formally outlined the concept of sustainability as "meeting the needs of the present without compromising the ability of future generations to meet their own needs"(*1*). Hence the concept is integrated with the development of the infrastructure. Sustainability access how the systems' performance impacts the environment, economy, and society. When the sustainability of a system is measured, all the components and the subsystems are considered simultaneously as they are closely related like a network (*2*). Especially for transportation infrastructure, sustainability has been a topic of concern for policymakers and planners as it is a major contributor to global air pollution, consumes the time of people in roadways, and is energy-intensive when providing mobility with vehicles (*3*). As Such, the transportation systems' performance creates a significant impact on the three components, i.e. environmental, social, and economic sustainability. However, the understanding and measurement of sustainability in transportation infrastructure systems have been described in many ways, also the quantifying indices vary substantially.

In addition, the transportation system requires to be more resilient in its service life to adopt any impact of sudden disruption by natural hazards, or any perturbation by external or internal events. The concept of resilience measures the system's functionality after any disruption as robustness, and recovery rate as rapidity. Moreover, the systems' alternative options for adopting the situation and available resource for that is also included in resiliency assessment. With the increasing risk of transportation system's vulnerability to disruption, design philosophies are adopting resilience indicators in the core principle. As such, the concept is integrated into all the construction phases i.e. planning, design, implementation, operation, maintenance of transportation infrastructure systems. Yet the quantification of resilience also has different philosophical perspectives and less standardized approaches in the literature.

It is crucial to consider both sustainability and resilience in transportation infrastructure systems at the same time. The implementation of sustainability in transportation systems will maximize resource utilization without compromising performance, and resiliency will make the systems more adaptive to disruption. This emerging concept of unification of sustainability and resilience is going to make a paradigm shift to new philosophies in each stage of transportation infrastructure construction phases. Moreover, the operation and performance are likely to improve with the intrusion of this new concept.

The critical review of existing literature on the unification of sustainability and resilience shows that some theoretical guidelines have been proposed but there is a lack of an effective single framework for their integration and quantification covering all effective indicators (*4–9*). Such a framework is necessary for their incorporation into transportation systems' construction stages. But unifying and improving sustainability and resilience does not always result in the enhancement of infrastructure because the same judgment which is made based on sustainability assessment may weaken the resilience of the infrastructure and vice versa (*4*). Moreover, sustainability is a more future-looking holistic approximation of possibility-based decisions choice, and resilience measures the immediate system's response to any perturbation. Therefore, it is a challenge for future researchers to quantify these two distant assessment approaches in the same unification framework.

This study reviewed existing literature to explore the concept of sustainability and resilience. Moreover, this study explored literature that focused on their quantification and integration framework. The main knowledge gap in the literature is found as the lack of a single unified framework that considers all the transportation infrastructure systems related indicators and quantifies sustainability and resilience simultaneously. Therefore, this research aims at finding the best describing indicators of resilience and sustainability of transportation infrastructure systems and unifying them in a single framework. The research is organized in the following steps:

1. Extensive literature search to identify the definitions, measuring indices, and relationships between sustainability and resilience with transportation infrastructure systems.

2. Finding the specific indices and identifying the relative interdependencies, tradeoffs with sustainability and resilience concepts.

The paper is organized in the following way. The second section will discuss the sustainability framework in transportation infrastructure systems. The third section will briefly delineate the outlines of resilience in the transportation network and its assessment. The fourth section will elaborate on the specific indices to access sustainability and resilience in the same unit and their relationships. In the final section, the article will be concluded by discussing the key findings and future scope of research.

**THE CONCEPT OF SUSTAINABILITY**

The concept of sustainable development was first formally delineated by the Brundtland Commission in 1987(*10*) and subsequently the concept is adopted in many works. In the year 2007, Johnston et al. estimated that some three hundred definitions of sustainability and sustainable development exist in different areas of literature (*11*). Still the lack of a standard definition and quantifying approach of sustainability for transportation infrastructure system exists (*12*). But some definitions paves the way to capture the overall idea, like in urban transportation sustainability is defined as "providing mobility with little or no harmful impact on health and environment and is providing mobility that ensures economic prosperity at no danger of depleting limited natural resources" (*13*). The very concept of sustainability focuses on the future development and sets goals to attain certain level of transportation system's performance based on heuristic indicators that may or may not change over time. This characteristic makes the measurement approach to a high sophistication level requiring the involvement of all five phases of project implementation program. Transportation agencies related to management, design, policy, and research believe the idea that sustainability must be measured from three equally weighted perspectives: economic, environmental, and social sustainability (figure 1) (*12*). As such, the indicators to measure the system's sustainability should comprise these three foundations, but in reality different agencies adopt different approaches to measure them. On the other hand, Zapata et al. think that, instead of finding the definition, the sustainability can be conceptualized as a body of knowledge and classify the study approach in stages related to environmental, social and ecological systems (*14*) as are highly affected by the transportation infrastructures. Though some studies indicate methodologies to measure sustainability considering the baseline of three foundations with relevant quantifiable indices (*12*, *15–18*), defining a generalized effective single framework of sustainable transportation systems still remains as a big challenge.

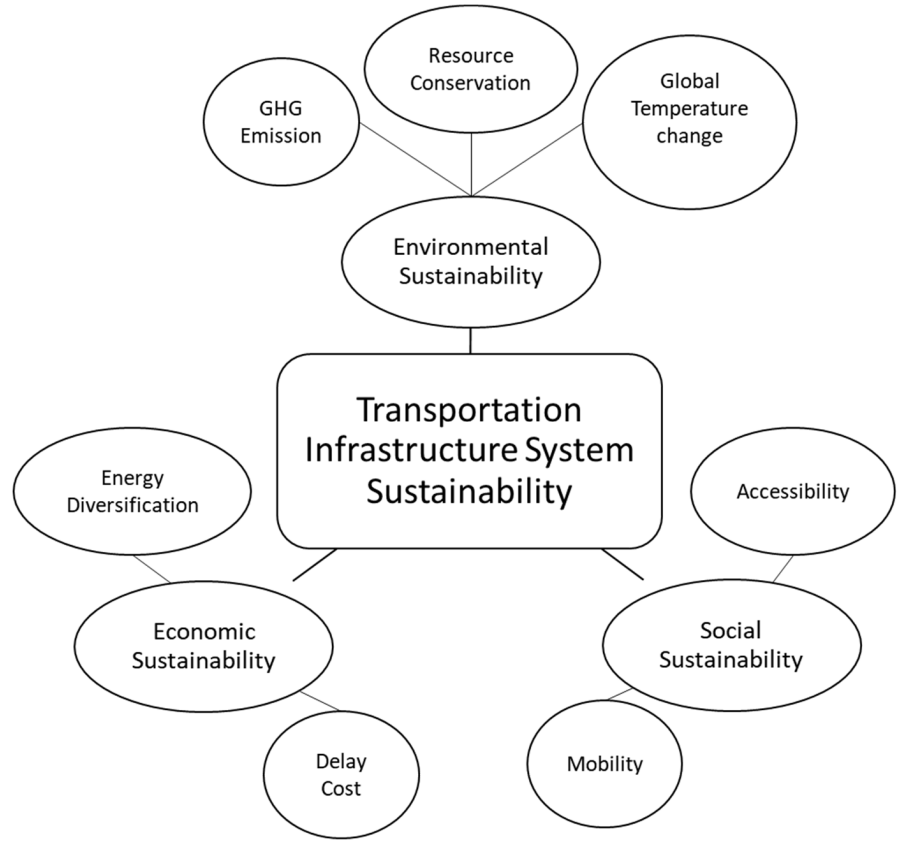

Figure 1: The Transportation Infrastructure Systems Sustainability Framework

**THE SUSTAINABILITY MEASURES**

Sustainability measure in transportation infrastructure must include all the subsystems otherwise the findings would become inconclusive to be meaningful. In this regards, Mahdinia et al. proposed an algorithm framework using principal component analysis to measure relative sustainability using 89 indicators (*18*). The author studied all the states of the United States and ranked them using relative sustainability indicators. Labib et al. studied the traffic network in Dhaka City and estimated emission to bio-capacity of the city, hence measuring the ecological footprint (*17*). Kennedy measures and compares the sustainability of public and private transportation systems in the Greater Toronto Area (GTA) based on an economic, social, and environmental framework (*16*). For economic analysis, cost to benefit analysis of the transportation components is used, like, the passenger kilometers travel is used to measure trip cost; for environmental sustainability, emission of different gas per capita is used; for environmental sustainability, level-of-service, insurance cost, accident data, and employment etc. are used as indicators. The outcome of the research is the public transportation system contributes more than private modes to sustainability but argued that private transport is essential for the system requiring policy-related approaches to make it more sustainable. Jeon et al. estimate sustainability indicated as an index for the Atlanta Metropolitan area and presented it as a decision support tool for future development (*15*). The study proposed the 'Sustainability triangle' which provides a good visual of the overall picture of the systems and helps to make decisions easier (figure 2). In the same study the overall sustainability is measured using environmental, social, and economic frameworks having twenty-eight parameters with customized weights. Al-Atawi et al. ranked sustainability parameters using Analytic Hierarchy Process Sustainability Compound Index for the city of Tabuk in Saudi Arabia (*19*). The parameters are weighted based on the users' and policy makers' opinions about the transportation facilities using travel behavior surveys. Life Cycle Cost analysis is also a very popular tool to measure sustainability (*4*). So, all these approaches prove that the transportation infrastructure system's assessment from a sustainability perspective is a time demanding research question and solving this puzzle is particularly difficult due to the lack of some standard platform.

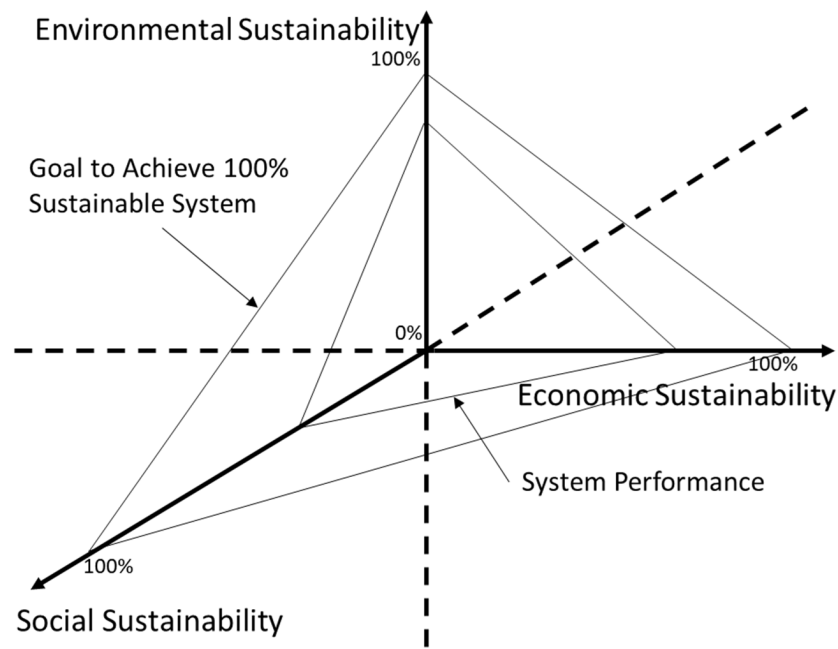

Figure 2: Sustainability Triangle for Decision Support (*15*)

**THE CONCEPT OF RESILIENCE**

The concept of resilience is a broad idea that measures any system's ability to adapt to any disruption, especially by natural hazards or unexpected incidents. System resilience depends on its flexibility, elasticity, and recovery, i.e. the more flexible and elastic the system, the less the recovery time hence represents more resilient. In an urban environment, the acceleration of development requires adaptation to rapid change, especially the change brought about by sudden disaster, which is very much visible due to the COVID-19 pandemic. In transportation infrastructure, as a part of the urban system, resilience is defined as the ability of the transportation network to absorb shock by man-made or natural disaster, provide mobility after the incident and restore the travel time to facilitate the increased traffic demand (*20*). The resiliency triangle in figure 3 explains the shock absorbing procedure of the system (*4*, *21*, *22*). After the disruptive incident takes place, the transportation system lost its functionality, and the remaining ones e.g., the remaining serviceability of the system is robustness of the system. The systems' pace to recover termed as rapidity which indicates how the system manage the excess demand with reduced capacity. The way rapidity gains to the steady state condition are termed as redundancy, e.g., alternative plans like new route choice or mode choice. To execute all these steps, a system needs to have resources, which is another measure of a system's resilience. The challenges of becoming resilient will require unique solutions to all four components as a whole for rapid transformation of the network within a short time in the implementation stages (*5*). So resilience is perceived as a multidisciplinary (*23*) framework, where the systems' adaptive capability depends upon the response of integrated sub-systems (*7*). Henry et al. thinks there is no consistent quantitative approach to measure resilience as the very concept is inconsistent to measure in different scenarios, also the limited approaches are confined to the discipline where it is measured (*21*). Resilience in transportation infrastructure accesses the systems' present functionality in case of emergency, and the systems' behavior is likely to be different case to case. But the methodology for the assessment still requires standard platform. Hence the methods of systems' resiliency assessment require to answer lots of questions like how to improve the four dimensions of the resilience model, what are the relationships and tradeoffs of these resilience dimensions, etc.

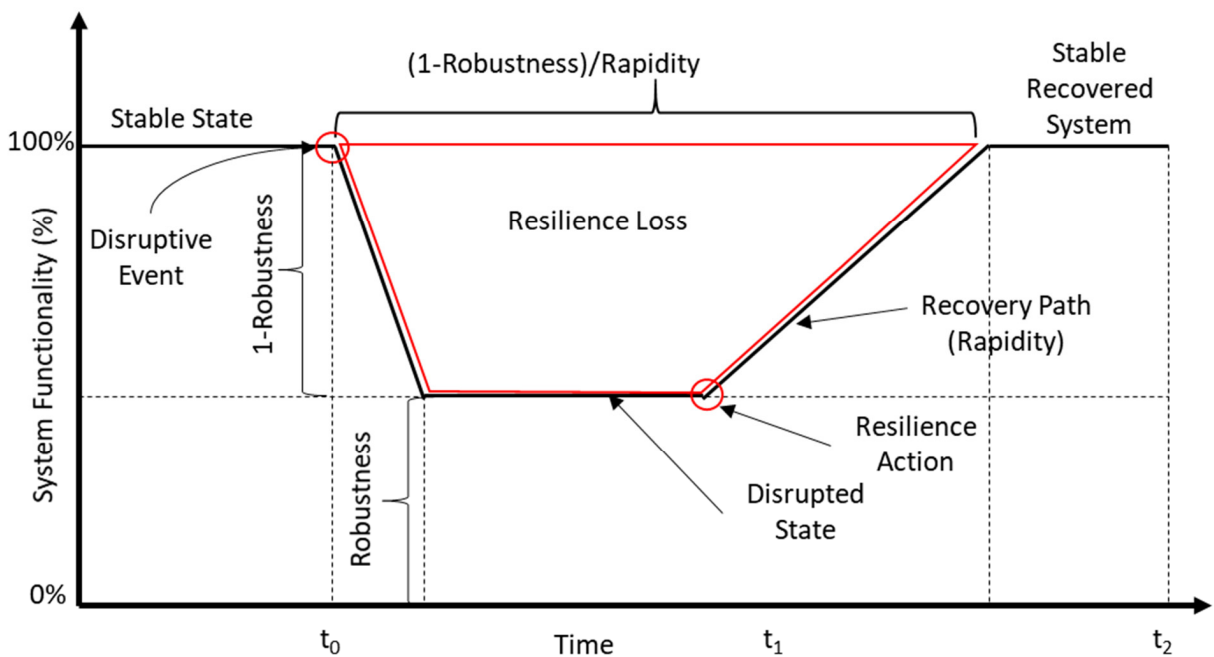

Figure 3: System Functionality and Concept of Resilience Measure (*4*, *21*, *22*).

## THE RESILIENCE MEASURE

Transportation infrastructure consists of lots of subsystems which eventually can be categorized into physical components and entities. Physical components include vehicles, equipment, power systems, fuel, control communication, and location systems. The entities that benefit to and from transportation infrastructure include government, general people, the supply industry, the financial community, and the competitors (*24*). Therefore, resilience study includes all these components and requires to be measured for better approximation. System's resilience assessment approach is divided into two major categories: qualitative and quantitative approach (*22*), wherein qualitative approach conceptual frameworks are developed to organize the strategies and design principles, and in quantitative approach, specific indices are measured to identify the relative index of resilience. Using a resilience triangle to determine resilience is a simple deterministic approach to quantification (*4*, *21*, *22*, *25*). Moreover time dependent recovery resilient metric otherwise known as dynamic resilience is used to assess the system's functionality and later normalize with cost function to evaluate the performance (*21*, *26*). Omer et al., on the other hand, develops a network of transportation infrastructure as well as the agencies and calculate the change in closeness centrality (CC) for network disruption (*27*). They used social network analysis for CC calculation and identify resilience from this indicator. Some other researchers adopt the same social network analysis to capture the evacuation, preparedness of community networks and measure the resilience of the transportation systems (*28–30*). Lots of other deterministic and probabilistic models are found in the literature (*22*). But an effective single standard framework or quantification procedure is absent in state-of-the-art transportation infrastructure resilience study, which is an open field to explore in future.

**SUSTAINABILITY AND RESILIENCE UNIFICATION FRAMEWORK**

Comparing the definitions of resilience and sustainability available in the literature, it is found that they have very different approaches to measure same performance of the system. Sustainability is a more inclined to work with present policies and goal for future performance estimation while resilience calculates the immediate systems' response to any extreme condition. But the bottom line is, in both cases the methodologies access the systems' performance. Moreover, the dimensions used in both concepts share common environmental, social, and economic indicators. Also, organizational reform, change in policy-related issues, adaptive design considerations are some of the common platforms where the outcomes of resilience and sustainability analysis are being used. Sustainability analysis is more expensive which requires a framework with a large number of designs, drawings, calculations, materials, implementations are taken into considerations and quantify the indices for future use. This heuristic analysis is only an estimation and may change over time. From the beginning of development of resilience study, it is quantitative in nature, and easily indicates the system performance based on numeric values. So, the analysis results are easily interpreted and integrated with the design, implementation, maintenance, and repair phases, while sustainability is difficult to adopt for many policymakers and transportation managers. One very distinguishable fact is that resilience is measured in a large spatial scale involving the communities. But sustainability is measured small spatial scale where the performance is confined to a specific project (*4*). However, sustainability related reform of policies, and philosophies are expected to strengthen the resiliency of the transportation systems, but the in-depth relationships and tradeoffs between them are still unknown (*5*).

From in-depth analysis of literature, the authors found some studies where sustainability and resilience are quantified using the similar indices. Vishnu et al. proposed a methodology to combining sustainability and resilience in the same framework and used travel time, distance, repair cost, repair $CO_2$, travel $CO_2$, and missing travel as the indicators. They also apply the methodology to calculate sustainability and resilience for Memphis Metropolitan Statistical Area. But the indicators seem too little to combine both sustainability and resilience. Roostaie and Nawari think that a unified framework may not be useful for projects rather the framework changes based on the nature of the project (*31*). However it is still difficult to measure all related indices in real terms even different frameworks for different projects are chosen(*32*). Xu et al. develop ideas of how resilience contributes to sustainability and how resilient systems can be sustainable. He chooses the specific field of social and ecological sustainability and reviews how to identify indicators and measuring criterion. Moreover, the study attempts to incorporate resilience into sustainability in an adaptive way (*33*), but specific guidelines or indicators are not articulated clearly.

The present study is confined to the transportation infrastructure systems and from extensive literature search, some common indicators are identified where the indicators of sustainability and resilience merge express each other. The indicators are summarized in table 1. It is identified that all different indicators discussed in sustainability and resilience can be expressed in terms of similar units as shown in figure 4 which will be the fundamental framework for this study. It converges from two directions: from sustainability and resilience. The main consideration for environmental sustainability is the emission from vehicles, which contributes to generating Green House Gas (GHG) in the atmosphere, impacting the global climate change. This change in phenomenon leads to extreme weather conditions, hence natural disasters become more frequent than before, requiring systems' resilience to be improved. Consequently, the vehicle's speed and overall system's delay contribute to the emission as well, therefore becoming an important indicator for sustainability. On the other hand, the system's capacity and resources are indicated by average speed and travel time which is a critical part of resilience investigation. Availability of different sources of energy resources increases resilience and at the same time reducing too much dependency on biodiesel and diversified sources affect sustainability. People spend a significant amount of time in transportation network for different purposes, as such mobility and accessibility are good performance indicator for this situation. Hence both sustainability and resilience of the system are affected by the systems' mobility and accessibility. A thorough analysis of these indicators in the next section will reveal the path to discover the relationships, tradeoffs between them.

**Table 1: Summary of Indices Discussed under Transportation Infrastructure Systems Sustainability and Resilience**

| Indicators | Units | Definition in Sustainability | Definition in Resilience |
|---|---|---|---|
| Emission of Green House Gas (GHG) | unit mass of GHG | Increased GHG contributes to global climate change | Increased GHS indicates the system has more delays, and less vehicle speed- meaning reduction of systems' performance |
| Average speed | distance over time | Speed within limit depending upon the mode of transport shows reduce emission | Increased speed means more capacity of the system |
| Change in ambient temperature | $^0f/^0c$ | The ambient temperature within a boundary reduces emission and fuel consumption | Change in temperature due to global warming leads to change in behavior of systems' elements and more frequent exposure to disaster |
| Energy source and diversification | gallon/liter per freeway miles | More energy source meaning less dependency on fossil-based fuel | Fuel diversification does not but its supply chain and availability are important for systems' resilience |
| Vehicle delay | unit time | Important indicator measuring systems' performance in environmental, social, and economic sustainability | Indicates systems' capacity and flexibility to absorb any external shock |
| Mobility | passenger/ton-miles | Measures systems' efficiency and time spent on travel | Indicated availability of systems' resources |
| Accessibility | relative quantification based on time, money, comfort, and risk | Provide systems' efficiency and availability of alternative mode choice | Measures how quickly the system can gain recovery with alternative resources |

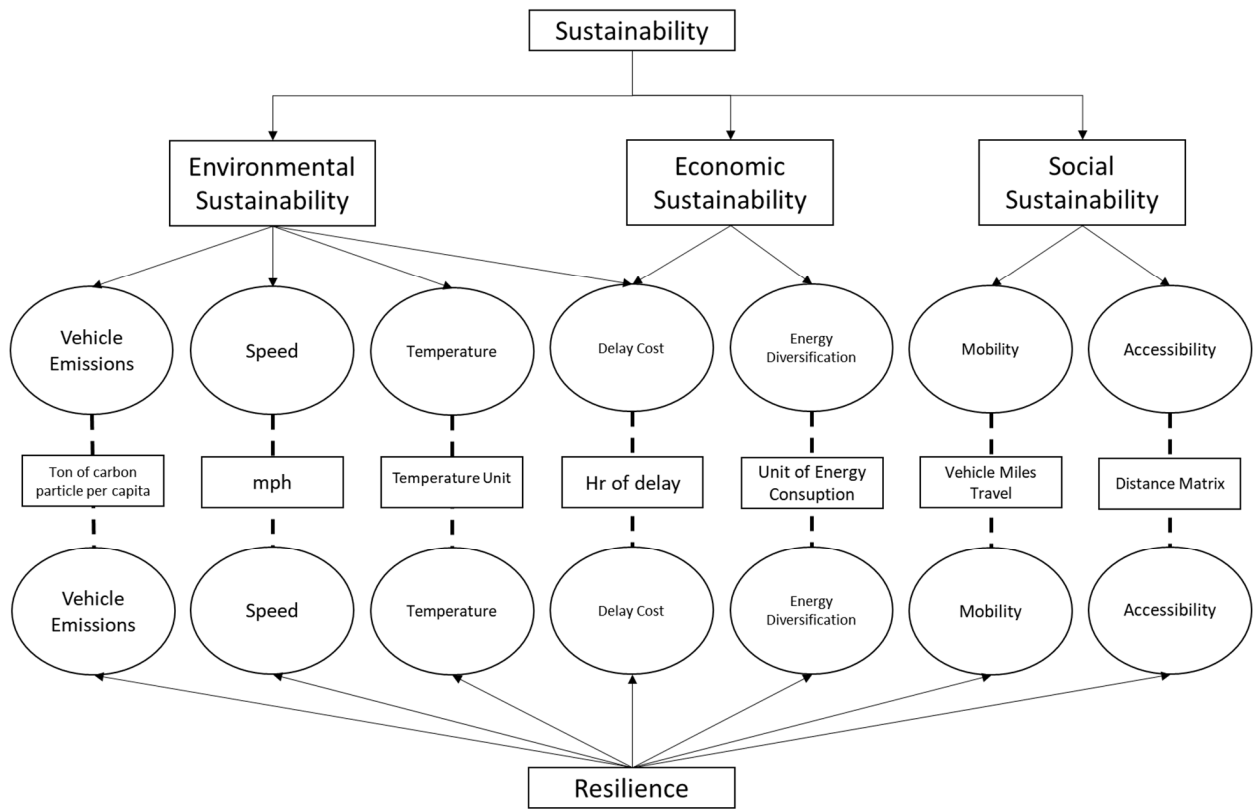

**Figure 4: Unified framework for sustainable and resilient transportation system**

**VEHICLE EMISSION**

Transportation is one of the major sources of environmental pollution including the emission of GHG, CO, $NO_x$, particulate matter, and volatile materials. Three main components of the transportation industry are contributing to this; vehicle, fuel, and mobility (*34*). In 2016, transportation infrastructure systems contributed more pollution than that of the power sector in the USA (*35*). Mahdinia et al. make a noble contribution in measuring sustainability in different states of the USA (*18*). It is estimated that increased emission from the transportation sector contributes to the reduction of the system sustainability (*36*). This is because the emission contributes to global climate change, which in turn affects the ecosystem in the future.

However, no direct relation between emission and resilience is not found in the literature, but the dependencies can be proven based on some other criterion. Transportation system resilience has a direct relationship with vehicle speed, delay which are responsible for emissions in air. As such, the relationship can be established as with increased delay, and reduced speed, system tend to produce more emission, also becomes less resilient. Hence emission modeling is a great indicator to be used to measure both resilience and sustainability in the same unit. The relationship can be represented in figure 5 schematic diagram.

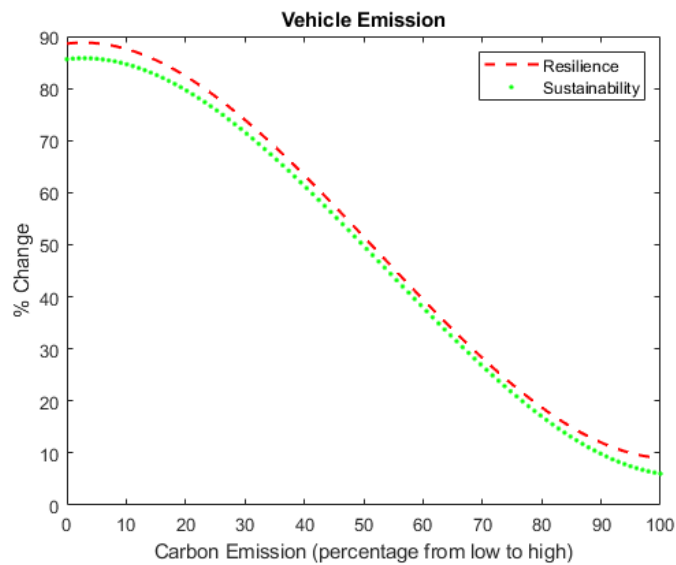

**Figure 5: Change of transportation system resilience and sustainability with change in vehicle emission (relative quantification)**

**AVERAGE SPEED CHARACTERISTICS**

Speed is an important indicator that can be used to quantify both the sustainability and resilience of transportation systems directly. Speed is the property of the vehicle and one of the performance parameters of the system. The network performs well when the average speed of a vehicle is higher, which indicates the lowest travel time and delay. But the speed of a vehicle has a direct relationship with emission. The emission is the lowest when the speed of the vehicle is medium, which is typically 30 to 50 mph, considering other parameters remain unchanged (*3, 37*).

At higher speeds, vehicle engines tend to operate at higher intensity resulting in incomplete combustion of fuel, leading to higher carbon emission e.g., high carbon emission at the freeways. Also, when the vehicle is operating at a lower speed or the network is heavily congested, the emission is higher due to higher travel time. The same situation is also true for an electrified vehicle, as they use higher energy at low and higher speeds, and this higher energy tradeoff higher emission at the electricity power plants. So, from a sustainability point of view, the transportation network is more sustainable when the speed is in the middle range.

The scenario is different in network resilience. In case of a disruption in the network, manmade or natural hazards, the network is likely to lose some of its performance and based on the system's resiliency, the network will take some time to recover. Within this time, some connections may be disrupted, and the travel pattern is likely to change, which results in increased travel time and reduced speed. It is found that the more average speed the network has, the more resilient the system is, as an indication of the infrastructure capacity. This resiliency of the network and average speed relationship has been tested in Paris Mass Railway Transportation System (*38*), and the author found that the system is quite resilient to maintain the part of passenger flow at lover operating speed.

Thus, the average speed is a good performance indicator for both resilience and sustainability of any transportation system, and the extent of this relationship has been shown schematically in figure 6. But it is not known at which speed, the system will be optimized for both resiliency and sustainability. Finding the tipping point will be a great research question for future exploration.

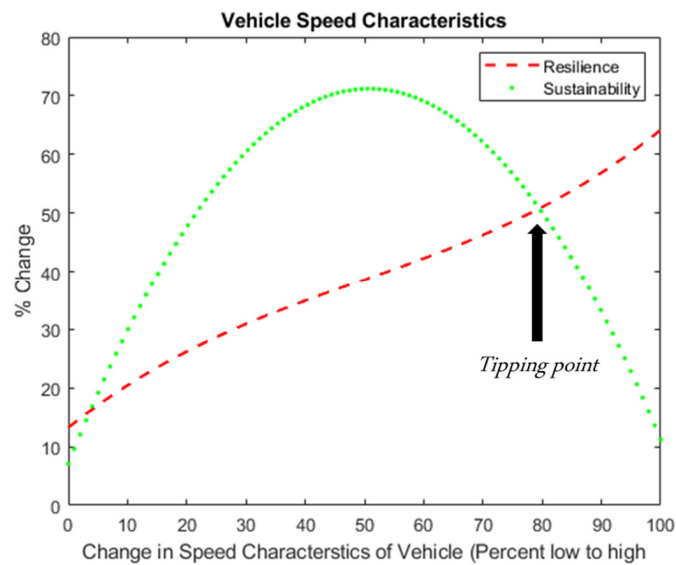

**Figure 6: Change of transportation system resilience and sustainability with change in vehicle speed characteristics (relative quantification)**

**CHANGE IN AMBIENT TEMPERATURE**

Ambient temperature is a good indicator to measure the environmental sustainability of the system, as it affects the evaporative emission. At higher temperatures, evaporative emissions increase during engine operation and rest. Emission is also affected in cold weather, as the engine efficiency is likely to decrease. At lower efficiency, some portion of the fuel is likely to remain unburnt, resulting in hydrocarbon emission (*3*, *37*). It is found that during summer and winter, emissions from the transportation sector increased (*39*). Moreover, at a higher and lower temperature, vehicles use extra energy for cooling and heating resulting in more fuel consumption. So, in extreme weather conditions, i.e., too much higher, or lower than average temperature, the system's sustainability is likely to decrease.

On the other hand, higher atmospheric temperature is a driving force for natural hazards, and the transportation infrastructure is prone to perturbation more frequently. Increased extreme warm or extreme cold weather increases the probability of pavement deterioration, changes the properties of pavement materials, and changes the behavior of expansion joints of bridges, etc. These phenomena may increase the maintenance of transportation network elements, resulting in frequent maintenance disruption (*40*). So, the temperature is an important indicator for quantifying the resilience of the system. It is found that the system performs more efficiently when the temperature is in the medium range(*41*).

But it is unknown the range of temperature at which the system starts losing its resilience and sustainability index, which can be a great future research. The heuristic relationship between temperature, sustainability, resilience can be expressed as in figure 7.

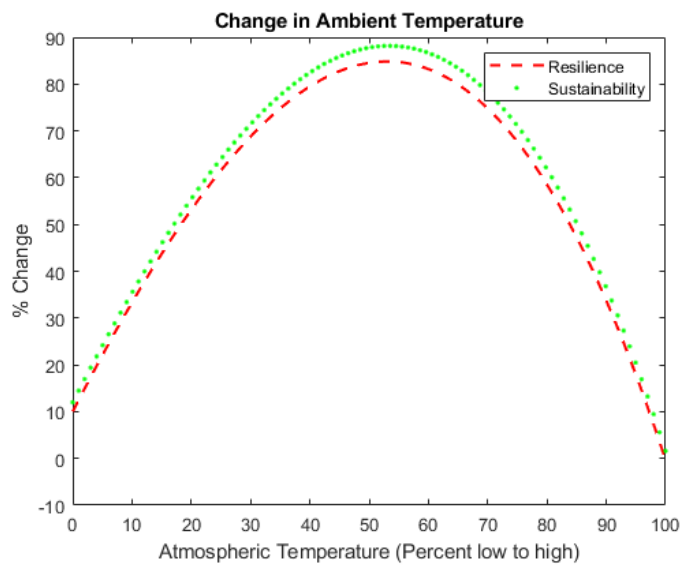

**Figure 7: Change of transportation system resilience and sustainability with change in atmospheric temperature (relative quantification)**

**ENERGY SOURCE AND DIVERSIFICATION**

Diversity of energy source is considered as security to uninterrupted supply and distribution of energy. Too much dependence on fossil-based energy is making the transportation system vulnerable for future performance (*37*). It is found in the literature that the diversified and in some cases locally available energy systems reduce too much pressure on a single source and improves environmental and economic sustainability in the long run (*42*). Especially in the transportation sector, electricity is considered to be the best alternative to bio-diesel, while sources of electricity generation are very much diversified (*43*).

The availability and the diversity of fuel sources have a direct effect on transportation system resiliency. Fuel supply is more resilient to any disruption or disaster if the source is diverse, and alternative resources are locally available. The availability can be improved through local storage, secure supply chain, availability of energy conversion with efficiency (*44*). But it is important to signify the interdependencies of different fuel sources and usability during any disaster, too much diversification can become a problem for the system functionality. In case of failure of one source may disable the users if the proper conversion system is unavailable. So, it is likely to affect the resourcefulness of the system, hence reducing the resilience at certain points.

Consequently, diversified sources of fuel can be a great indicator where a system's resilience and sustainability can be quantified together. But the research question is at which point the tradeoff between sustainability and resilience will likely take place, which is still unknown to the society of researchers. The relationship between sustainability and resilience can be visualized as in figure 8.

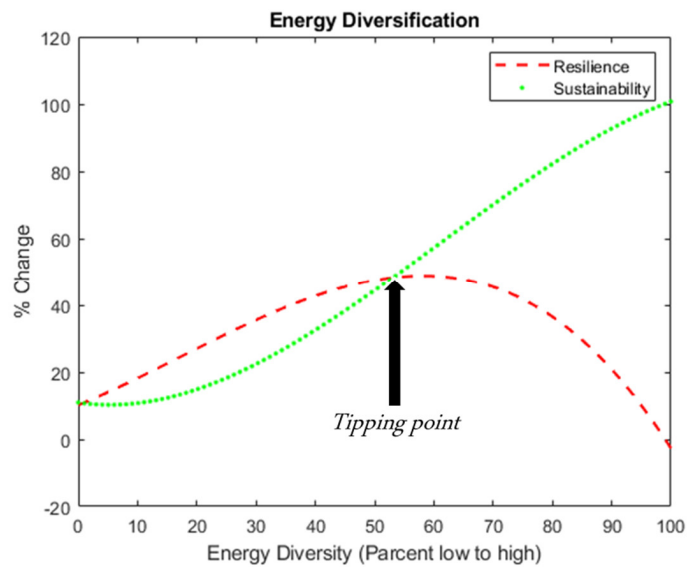

**Figure 8: Change of transportation system resilience and sustainability with change in energy diversification (relative quantification)**

## DELAY

One of the important parameters to measure network performance is travel time. Delay is explained when the system's travel time is greater than the average. In transportation infrastructure systems delay is associated with huge costs which even affect the overall GDP, indicating loss of sustainability and resilience. With an increased delay, the system will experience higher emission, and travel time, that reflects the wastage of resource, hence reduces sustainability measure (*45, 46*). The same scenario is observed in network resiliency. Under extreme conditions or any stress in the network, system performance is likely to be affected and travel time increases. In this situation, the less the resilience of the network, the more delay it will suffer (*47, 48*). Delay has a direct effect on economic, environmental, and social sustainability, also affecting resilience. The system is likely to lose its sustainability and resilient index with increase in delay, but the rate of decrease in a unifying framework is an unanswered question. The dependencies of resilience and sustainability on delay can be shown in figure 9.

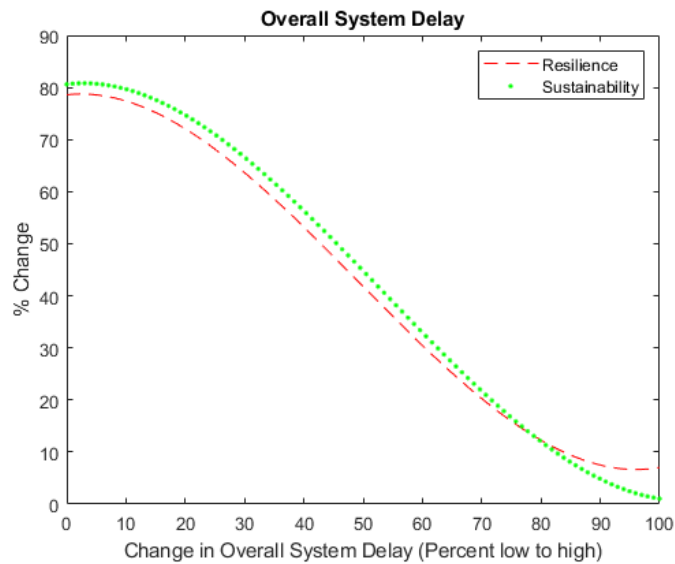

**Figure 9: Change of transportation system resilience and sustainability with change in system's delay (relative quantification)**

## MOBILITY

Mobility is the movement of people within the transportation network and is directly linked with the sustainability and resilience of the system. Due to scarcity of land and limited capacity of links, the demand for mobility is very high and always considered in the planning goal of the urban infrastructure systems (*49*). The availability of increased mobility is a good indicator of urban sustainability, i.e. especially from a social perspective as it indicates the systems' efficiency to accommodate the demand, hence reducing time in the road or other mode of transport. But this does not limit to increasing mobility of a specific mode of transport rather sustainability increases with the shared mobility, e.g. public transport, walking, cycling, and more localized travel pattern (*50*).

The same trend, as observed in sustainability, is true for a system's resilience. The consequence is especially visible during the COVID-19 pandemic. The transportation system is on suspension in many countries in the world. But people choose other modes of communication to make the system operational. From a resilience point of view, physical mobility is totally disrupted but online communications, i.e. mobility in other forms, keep the system functional (*51, 52*). But in general, resilience is likely to be degraded with the reduction of mobility. In this context, the relationship between sustainability, resilience, and mobility can be expressed as shown in figure 10.

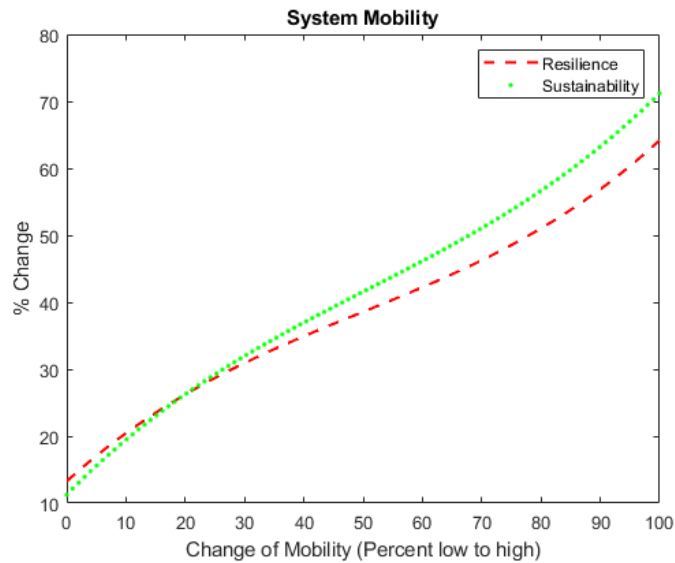

**Figure 10: Change of transportation system resilience and sustainability with change in system mobility (relative quantification)**

**ACCESSIBILITY:**

Traditional transportation planning is focused on mobility-based indicators, but mobility is a means to get access to people's requirements to make the trip. Now the concept is increasing accessibility within the urban system which is not always the physical movement, rather another mode of transportation (*53*). Accessibility can be increased in the urban system by increasing different means of communication, therefore reducing physical movement to some extent. So, accessibility adds sustainability to the system, because the emission and delay are likely to be reduced with increased accessibility (*54*). So, the industries, like tech-industries, are trying to implement the concept of a smart city, where the employees can do their job without making significant movement.

On the other hand, unexpected network perturbation will reduce the transportation network capacity. If the proper recovery plan is not in action, there is a possibility of a loss of people's movement within the system, leading to loss of resilience (*55, 56*). Henceforth accessibility is a great indicator to be used as a part of systems' resilience estimation. In general, the relationship between them is, the more accessibility the system has, the more resilient it is. Therefore, the measure of accessibility is a good indicator important for both sustainability and resilience measures of the transportation system, and the dependencies can be shown as in figure 11.

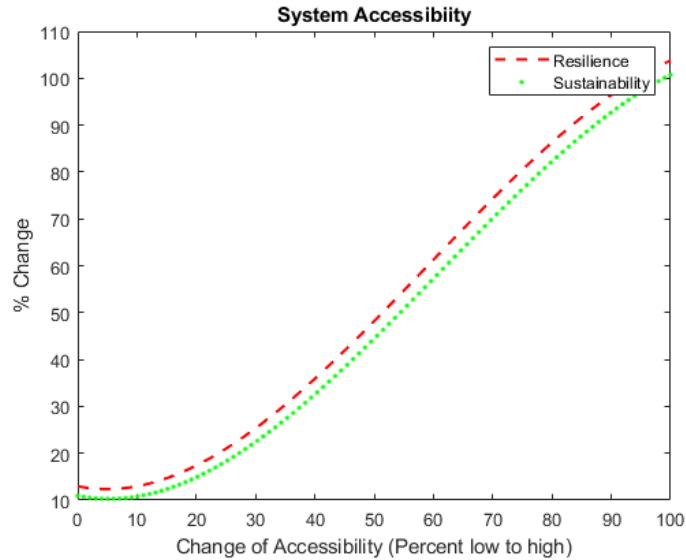

**Figure 11: Change of transportation system resilience and sustainability with change in system's accessibility (relative quantification)**

## CONCLUDING REMARKS AND SCOPE FOR FUTURE RESEARCH

Sustainability measures the impacts transportation infrastructure systems' performance on environmental, economic, and social systems. It deals with policy related issues and integrated in planning, design, implementation, operation, and maintenance phases of the system to measure the impacts. On the contrary, transportation systems' resilience evaluates the ability of the system to absorb any shock induced by natural disaster or any unexpected events. Resilience measures systems' functionality after the disruption i.e. robustness and, rate of gain of functionality as rapidity. Moreover, resilience also quantifies the resourcefulness and redundancy of the system. In order to make the transportation infrastructure system more durable, the considerations for both sustainability and resilience are equally important in all the construction phases otherwise the systems are likely to suffer in their design life. As such unified thinking of sustainability and resilience will benefit the transportation systems' performance for longer time. Though some unified resilience and sustainability frameworks are available in the literature, most of them are in theatrical form and lack complete guidance to measure both concepts simultaneously. Some studies quantify sustainability and resilience but used limited indicators which partially represent the system. In contrast, the same indicator reinforces one and undermines another in certain thresholds, which directs us to evaluate the system from all directions. Considering all these bottlenecks and possibilities this study proposed an effective single unifying framework that can measure the transportation systems' performance simultaneously from sustainability and resilience perspective. The proposed unifying framework address seven critical indicators including emission, speed, temperature, energy consumption, delay, mobility, and accessibility to perform the systems' assessment. This study also identifies the relationships, interdependencies, and tradeoffs between sustainability and resilience based on these indices. Moreover, some of the indicators have some tolerance limit beyond which they show inverse behavior in estimating sustainability and resilience, like the vehicle speed characteristics which improves sustainability when ranging from 30 to 50 mph but improves resilience beyond that limit. Similarly, too much energy diversification adds more points to systems' sustainability but make less resilient.

These characteristics warrants for immediate research in the field of resilience and sustainability. For future research in areas involving unification of sustainability and resilience, few possible research questions might be:

- What are the tipping points where the transportation systems can achieve maximum resiliency and sustainability?
- What is the scaling effect on the systems' unified assessment?
- How will the global community adopt the same and progressive design philosophy considering the unified sustainability and resilience?
- What innovative modifications are requires shifting from traditional to new design concepts?
- How the behavior of user change with new adaptation?

Some new research directions can be designed to calibrate and validate the framework which includes working with primary data from case studies. In order to evaluate the scaling effect, multiple case studies are required. In case of adaptation strategies, it is important to get the views of uses and providers through statistical analysis. The new unification approach is likely to change the traditional philosophies in design and adopt new concepts. To predict the probable modifications requires extensive case studies are required, not limited to space and time, rather beyond the boundary of countries.

**AUTHOR CONTRIBUTION STATEMENT**

Kays, Sadri                                                                                                          20

Kays, Sadri                                                                                                                                                21